\documentclass[preprintnumbers,superscriptaddress,showpacs,pra]{revtex4}
\usepackage{amssymb,amsmath}
\usepackage{graphicx}
\usepackage{amsmath}
\usepackage{amsfonts}

\input{tcilatex}
\begin{document}

\preprint{}
\title[geometric momentum and geometric potential]{Can the canonical
quantization be accomplished within the intrinsic geometry?}
\author{D. M. Xun}
\affiliation{School for Theoretical Physics, and Department of Applied Physics, Hunan
University, Changsha, 410082, China}
\author{Q. H. Liu}
\email{quanhuiliu@gmail.com}
\affiliation{School for Theoretical Physics, and Department of Applied Physics, Hunan
University, Changsha, 410082, China}
\date{\today }

\begin{abstract}
For particles constrained on a curved surface, how to perform quantization
within Dirac's canonical quantization scheme is a long-standing problem. On
one hand, Dirac stressed that the Cartesian coordinate system has
fundamental importance in passing from the classical Hamiltonian to its
quantum mechanical form while preserving the classical algebraic structure
between positions, momenta and Hamiltonian to the extent possible. On the
other, on the curved surface, we have no exact Cartesian coordinate system
within intrinsic geometry. These two facts imply that the three-dimensional
Euclidean space in which the curved surface is embedded must be invoked
otherwise no proper canonical quantization is attainable. Since the minimum
surfaces, catenoid and helicoid studied in this paper, have vanishing mean
curvature, we explore whether the intrinsic geometry offers a proper
framework in which the quantum theory can be established in a
self-consistent way. Results show that it does for quantum motions on
catenoid and it does not for that on helicoid, but neither is compatible
with Schr\"{o}dinger theory. In contrast, in three-dimensional Euclidean
space, the geometric momentum and potential are then in agreement with those
given by the Schr\"{o}dinger theory.
\end{abstract}

\maketitle

\section{Introduction}

Recently, a quantum motion on the two-dimensional curved surface $\Sigma ^{2}
$ has attracted increasing attention. \cite%
{jpn1990,jpn1992,Golovnev,Golovnev2,SJ,jk,dacosta,CB,FC,liu07,liu11,japan1990,japan1992,japan1993,liu13-1,liu13-2,helicoid,catenoid,Szameit,onoe}
On one hand, if the quantum Hamiltonian is the same function of the
canonical coordinates and momenta in the quantum theory as in the classical
theory, the canonical quantization needs a well-defined Cartesian coordinate
system. \cite{dirac1,Schiff,Greiner,weinberg} On the other, for a curved
manifold, not only the global Cartesian coordinate system does not exist but
also the local one can be used approximately. In the differential geometry
for the surface $\Sigma ^{2}$, a complete description of $\Sigma ^{2}$ needs
a three-dimensional flat space $R^{3}$ in which so-called second fundamental
form can be then defined. \cite{diffgeom} Combining these two observations,
we are confident that a proper description for quantum motion on $\Sigma ^{2}
$ is not possible unless in the three-dimensional flat space $R^{3}$. As a
consequence, the geometric momentum is identified and introduced as a proper
description of the momentum for a particle constrained on the curved
surface. \cite{liu07,liu11} In contrast, the conventional formalism of
quantum mechanics is established within the framework of the intrinsic
geometry. In this paper, we utilize two minimum surfaces, catenoid and
helicoid, to further explore the relationship between geometry and
quantization.

Let us first recall elementary differential geometry for the two-dimensional
curved surface $\Sigma ^{2}$ that is embedded in the three-dimensional
Cartesian space $R^{3}$. The surface $\Sigma ^{2}$ is parameterized by $%
q^{\mu }\equiv (u,v)$ with $\mu $ running from $1$ to $2$, we have in
three-dimensional Cartesian coordinate system the positions $\mathbf{r\equiv
(}x(u,v),y(u,v),z(u,v)\mathbf{)}$, and normal vector $\mathbf{n=(}%
n_{x},n_{y},n_{z}\mathbf{)}\mathbf{\equiv }$ $\mathbf{r}_{u}\times \mathbf{r}%
_{v}/\left\vert \mathbf{r}_{u}\times \mathbf{r}_{v}\right\vert $ where $%
\mathbf{r}_{\mu }\equiv \partial \mathbf{r/}x^{\mu }$, and $\mathbf{r}^{\mu
}=g^{\mu \nu }\mathbf{r}_{\nu }=g^{\mu \nu }\partial \mathbf{r/}x^{\nu }$.
In whole of this paper, the Einstein summation convention that repeated
indices are implicitly summed over. At this point $\mathbf{r}$, we have two
geometric invariants, the mean curvature vector $M\mathbf{n}$ and the
gaussian curvature $K$ which characterizes, respectively, the extrinsic and
the intrinsic curvature. 

Next, let us recall\ a fact on the relation between the three-dimensional
Cartesian space $R^{3}$ and an effective quantum theory for the surface $%
\Sigma ^{2}$. The Schr\"{o}dinger equation is first formulated in $R^{3}$,
actually in a curved shell of an equal and finite thickness $\delta $ whose
intermediate surface coincides with the prescribed one $\Sigma ^{2}$ (or
equivalently, the particle moves within the thin layer of the same width $%
\delta $ due to a confining potential around the surface), and an effective
Schr\"{o}dinger equation on the curved surface $\Sigma ^{2}$ is second
derived by taking the squeezing limit $\delta \rightarrow 0$ to confine the
particle to the $\Sigma ^{2}$. \cite%
{jk,dacosta,CB,FC,liu07,liu11,japan1992,japan1993,helicoid,catenoid,Szameit,onoe}
It leads to unambiguous forms for the geometric momentum $\mathbf{p}$ \cite%
{liu07} and geometric kinetic energy that contains the geometric potential $%
V_{g}$ \cite{jk,dacosta}, which are given by, 
\begin{equation}
\mathbf{p}=-i\hbar (\mathbf{r}^{\mu }\partial _{\mu }+M\mathbf{n}),
\label{geom}
\end{equation}%
\begin{equation}
T\equiv -\frac{\hbar ^{2}}{2m}\Delta +V_{g}\text{, }V_{g}=-\frac{\hbar ^{2}}{%
2m}\left( M^{2}-K\right) ,  \label{geokin}
\end{equation}%
where $\mathbf{r}^{\mu }\partial _{\mu }$ is sometimes denoted as $\nabla
_{2}$ that is the gradient operator on a two-dimensional surface. \cite%
{diffgeom} Both the kinetic energy (\ref{geokin}) and momentum (\ref{geom})
are geometric invariants.

The presence of the geometric potential $V_{g}$ enriches our understanding
of the quantization procedure. For a quantum system that has a classical
analogue, we can no longer assume in general that, the quantum Hamiltonian
is the same function of the canonical coordinates and momenta in the quantum
theory as in the classical theory, even in the Cartesian coordinate system.
Moreover, the consistence between \textit{fundamental quantum conditions }%
and the equations of motion, i.e., the Ehrenfest theorem $[f,H]/(i\hbar
)=\{f,H\}_{D}$ for $f=$ $x_{{i}}$ and $p_{{i}}$, turns out to be
problematic, where $\{A{,B\}}_{D}$ is in general the Dirac bracket between
two variables $A${\ and }${B}$ for a system that has second-class
constraints, and reduces to the usual Poisson bracket when the system is
constraint-free. By the \textit{fundamental quantum conditions, }we mean
that the commutation relations $[A,B]$ between the coordinates $x_{\mu }$ ($%
\mu =1,2$) and momenta $p_{\mu }$, or $x_{{i}}$ ($i=1,2,3$) and $p_{{i}}$,
satisfy $[A,B]/(i\hbar )=\{A,B\}_{D}$, according to Dirac. \cite{dirac1} The
current procedure \cite{pauli,2005} quantizes both the generalized
coordinates/momenta ($x_{\mu }$, $p_{\nu }$) and the Cartesian ones ($x_{i}$%
, $p_{j}$) on an equal footing. \cite{dirac1,dirac2} This procedure differs
from the underlying idea in the thin layer method in\ the Schr\"{o}dinger
equation approach as outlined above, where $q^{\mu }\equiv (u,v)$ are purely
parameters while performing quantization of the Cartesian coordinates and
momenta ($\mathbf{x},\mathbf{p}$). 

Now, we propose a \textit{strengthened} version of the Dirac's \textit{%
canonical quantization} (SCQ) scheme as what follows. For a quantum system
that has a classical analogue, there are two categories of the \textit{%
fundamental quantum conditions.} The original ones belong to the first,
which is between the coordinates $x_{\mu }$ and momenta $p_{\mu }$ or $x_{{i}%
}$ and $p_{{i}}$. The second category is the commutation relations $[f,H]$,
where $H$ is the Hamiltonian and $f=x_{\mu }$ and $p_{\mu }$ or $f=$ $x_{{i}}
$ and $p_{{i}}$, which must also satisfy correspondence $[f,H]/(i\hbar
)=\{f,H\}_{D}$. In other words, the SCQ hypothesizes a \textit{simultaneous }%
quantization for positions, momenta, and Hamiltonian while\textit{\ }%
preserving the formal algebraic structure between them to the extent possible%
\textit{.} It is evident that, once the constraints are free, the second
category of the \textit{fundamental quantum conditions }is trivially
satisfied because they are nothing but the Ehrenfest theorem, as long as the
Cartesian coordinate system is used.

Notice that the Dirac's canonical quantization scheme should be examined on
the case-by-case basis. \cite{weinberg}\textit{\ }For particles constrained
on the minimum surfaces with $M=0$, momentum (\ref{geom}) and kinetic energy
(\ref{geokin}) assume their dependence on purely intrinsic geometric
quantity. Whether the intrinsic geometry offer a proper framework for the
canonical quantization scheme is then an interesting issue. Recently,
quantum motion on two minimum surfaces, catenoid \cite{catenoid} and
helicoid, \cite{helicoid} are investigated. In the present paper we take
also these two surfaces to see whether the quantum theory can be established
satisfactorily. Results turn out to be compatible with Dirac's remark that
only the Cartesian coordinate system is physically permissible while the
intrinsic geometry suffers from various problems. The organization of the
paper is as follows. In sections II and III, we study catenoid,
respectively, within purely intrinsic geometry and as a submanifold in $R^{3}
$. In sections IV and V, we study helicoid in similar manner. Section VI
briefly remarks and concludes this study. In the present study, no external
potential field presents without loss of generality.

\section{Dirac's theory of second-class constraints for a catenoid within
intrinsic geometry}

The catenoid is with two local coordinates $\theta\in\lbrack0,2\pi),\rho\in
R $,%
\begin{equation}
\mathbf{r}=(r\cosh\frac{\rho}{r}\cos\theta,r\cosh\frac{\rho}{r}\sin\theta
,\rho),\text{ }r>0,  \label{rr}
\end{equation}
where $r$ is the constraint parameter that will be set as $r=a\neq0$. In
this section, we will first give the classical mechanics for motion on the
catenoid within Dirac's theory of second-class constraints, and then turn
into quantum mechanics. In classical mechanics, the theory appears nothing
surprising, but after transition to quantum mechanics, it breaks agreement
with the Schr\"{o}dinger theory.

\subsection{Classical mechanical treatment}

The Lagrangian $L$ in the local coordinate system is, 
\begin{align}
L& =\frac{1}{2}m\left( \frac{1}{r^{2}}\left( r\cosh \frac{\rho }{r}-\rho
\sinh \frac{\rho }{r}\right) ^{2}\dot{r}^{2}+r^{2}\cosh ^{2}\frac{\rho }{r}%
\dot{\theta}^{2}+\frac{2}{r}\sinh \frac{\rho }{r}(r\cosh \frac{\rho }{r}%
-\rho \sinh \frac{\rho }{r})\dot{r}\dot{\rho}+\cosh ^{2}\frac{\rho }{r}\dot{%
\rho}^{2}\right)  \notag \\
& -\lambda (r-a),  \label{lag}
\end{align}%
where $\lambda $ is the Lagrangian multiplier enforcing the constrained of
motion on the surface, nevertheless, we treat the quantity $\lambda $ as an
additional dynamical variable. The Lagrangian is singular because it does
not contain the "velocity" $\dot{\lambda}$. Hence we need Dirac's theory,
which gives the canonical momenta conjugate to $r,\theta ,\rho $ and $%
\lambda $ in the following,%
\begin{align}
p_{r}& =\frac{\partial L}{\partial \dot{r}}=m\left( \frac{1}{r^{2}}\left(
r\cosh \frac{\rho }{r}-\rho \sinh \frac{\rho }{r}\right) ^{2}\dot{r}+\frac{1%
}{r}\sinh \frac{\rho }{r}\left( r\cosh \frac{\rho }{r}-\rho \sinh \frac{\rho 
}{r}\right) \dot{\rho}\right) , \\
p_{\theta }& =\frac{\partial L}{\partial \dot{\theta}}=mr^{2}\cosh ^{2}\frac{%
\rho }{r}\dot{\theta}, \\
p_{\varphi }& =\frac{\partial L}{\partial \dot{\varphi}}=m\left( \frac{1}{r}%
\sinh \frac{\rho }{r}\left( r\cosh \frac{\rho }{r}-\rho \sinh \frac{\rho }{r}%
\right) \dot{r}+\cosh ^{2}\frac{\rho }{r}\dot{\rho}\right) , \\
p_{\lambda }& =\frac{\partial L}{\partial \dot{\lambda}}=0.  \label{plamb}
\end{align}%
Eq. (\ref{plamb}) represents the primary constraint:%
\begin{equation}
\varphi _{1}\equiv p_{\lambda }\approx 0,  \label{prim}
\end{equation}%
hereafter symbol "$\approx $" implies a weak equality \cite{dirac2}. After
all calculations are finished, the weak equality takes back the strong one.
By the Legendre transformation, the primary Hamiltonian $H_{p}$ is, \cite%
{dirac2}%
\begin{equation}
H_{p}=\frac{r^{2}\cosh ^{2}\frac{\rho }{r}}{2m\left( r\cosh \frac{\rho }{r}%
-\rho \sinh \frac{\rho }{r}\right) ^{2}}p_{r}^{2}-\frac{r\sinh \frac{\rho }{r%
}}{m\left( r\cosh \frac{\rho }{r}-\rho \sinh \frac{\rho }{r}\right) ^{2}}%
p_{r}p_{\rho }+\frac{1}{2mr^{2}\cosh ^{2}\frac{\rho }{r}}p_{\theta }^{2}+%
\frac{1}{2m}p_{\rho }^{2}+\lambda \left( r-a\right) +up_{\lambda },
\label{hami}
\end{equation}%
where $u$ is also a Lagrangian multiplier guaranteeing that this Hamiltonian
is defined on the symplectic manifold. The secondary constraints (not
confusing with second-class constraints) are generated successively, then
determined by\ the conservation condition \cite{dirac2},%
\begin{equation}
\varphi _{i+1}\equiv \left\{ \varphi _{i},H_{p}\right\} \approx 0,\text{\ }%
(i=1,2,....),
\end{equation}%
where $\left\{ f,g\right\} $ is the Poisson bracket with $%
q_{1}=r,q_{2}=\theta ,q_{3}=\rho $, and $p_{1}=p_{r},p_{2}=p_{\theta
},p_{3}=p_{\rho }$, 
\begin{equation}
\left\{ f,g\right\} \equiv \frac{\partial f}{\partial q_{k}}\frac{\partial g%
}{\partial p_{k}}+\frac{\partial f}{\partial \lambda }\frac{\partial g}{%
\partial p_{\lambda }}-(\frac{\partial f}{\partial p_{k}}\frac{\partial g}{%
\partial q_{k}}+\frac{\partial f}{\partial p_{\lambda }}\frac{\partial g}{%
\partial \lambda }).  \label{possi}
\end{equation}%
The complete set of the secondary constraints is, 
\begin{align}
\varphi _{2}& \equiv \left\{ \varphi _{1},H_{p}\right\} =a-r\approx 0,
\label{db1} \\
\varphi _{3}& \equiv \left\{ \varphi _{2},H_{p}\right\} =r\left( -\frac{%
r\cosh ^{2}\frac{\rho }{r}}{m\left( r\cosh \frac{\rho }{r}-\rho \sinh \frac{%
\rho }{r}\right) ^{2}}p_{r}+\frac{\sinh \frac{\rho }{r}}{m\left( r\cosh 
\frac{\rho }{r}-\rho \sinh \frac{\rho }{r}\right) }p_{\rho }\right) \approx
0,  \label{db2} \\
\varphi _{4}& \equiv \left\{ \varphi _{3},H_{p}\right\}  \notag \\
& =\frac{r^{2}\cosh ^{2}\frac{\rho }{r}}{m\left( r\cosh \frac{\rho }{r}-\rho
\sinh \frac{\rho }{r}\right) ^{2}}\lambda +\frac{r^{2}\cosh \frac{\rho }{r}%
(rp_{\rho }-2\rho p_{r}+r\cosh \frac{2\rho }{r}p_{\rho }-(rp_{r}+\rho
p_{\rho })\sinh \frac{2\rho }{r})^{2}}{4m^{2}\left( r\cosh \frac{\rho }{r}%
-\rho \sinh \frac{\rho }{r}\right) ^{5}}  \notag \\
& -\frac{p_{\theta }^{2}}{m^{2}r^{2}\left( r\cosh \frac{\rho }{r}-\rho \sinh 
\frac{\rho }{r}\right) \cosh ^{3}\frac{\rho }{r}}\approx 0.  \label{thi}
\end{align}%
Eqs. (\ref{db1}) shows that on the surface of catenoid $r=a$, there is no
motion along the normal direction, while Eqs. (\ref{thi}) determines the
dynamical variable $\lambda $, and by the conservation condition of the
secondary constraint $\varphi _{4}$ (\ref{thi}), we can determine the
Lagrangian multipliers $u$.

The Dirac bracket instead of the Poisson bracket for two variables $A$ and $%
B $ is defined by,%
\begin{equation}
\left\{ A,B\right\} _{D}\equiv\left\{ A,B\right\} -\left\{ A,\varphi
_{\xi}\right\} C_{\xi\zeta}^{-1}\left\{ \varphi_{\zeta},B\right\} ,
\end{equation}
where the $4\times4$ matrix $C\equiv\left\{ C_{\xi\zeta}\right\} $ whose
elements are defined by $C_{\xi\zeta}\equiv\left\{
\varphi_{\xi},\varphi_{\zeta}\right\} $ with $\xi,\zeta=1,2,3,4$ from Eqs. (%
\ref{prim}) and (\ref{db1})-(\ref{thi}). The inverse matrix $C^{-1}$ is,%
\begin{equation}
C^{-1}=\left\{ 
\begin{array}{cccc}
0 & C_{12}^{-1} & C_{13}^{-1} & C_{14}^{-1} \\ 
-C_{12}^{-1} & 0 & C_{23}^{-1} & 0 \\ 
-C_{13}^{-1} & -C_{23}^{-1} & 0 & 0 \\ 
-C_{14}^{-1} & 0 & 0 & 0%
\end{array}
\right\} ,
\end{equation}
where%
\begin{align}
C_{12}^{-1} & =\frac{1}{ma^{6}\cosh^{4}\frac{\rho}{a}}(\left( -5p_{\theta
}^{2}+a^{2}p_{\rho}^{2}\right) \rho^{2}\frac{1}{\cosh^{4}\frac{\rho}{a}}%
+a\left( -p_{\theta}^{2}+a^{2}p_{\rho}^{2}\right) \left( 2a-\rho\tanh \frac{%
\rho}{a}\right)  \notag \\
& +\frac{1}{\cosh^{2}\frac{\rho}{a}}\left(
-a^{4}p_{\rho}^{2}+p_{\theta}^{2}\left( 5a^{2}+4\rho^{2}\right) +2a\left(
-5p_{\theta}^{2}+a^{2}p_{\rho }^{2}\right) \rho\tanh\frac{\rho}{a})\right) ),
\\
C_{13}^{-1} & =-\frac{p_{\rho}}{a^{3}\cosh^{5}\frac{\rho}{a}}\left( a\cosh%
\frac{\rho}{a}-\rho\sinh\frac{\rho}{a}\right) \left( 2\rho+a\sinh \frac{2\rho%
}{a}\right) , \\
C_{14}^{-1} & =-C_{23}^{-1}=\frac{m}{a^{2}}\left( a-\rho\tanh\frac{\rho}{a}%
\right) ^{2}.
\end{align}
Because of no motion along the normal direction, we need not analyze the
dynamics\ and kinematics of the normal direction. Thus, the generalized
positions $q^{\mu}$ $(=\theta,\rho)$ and momenta $p_{\nu}$ satisfy the
following Dirac brackets,%
\begin{equation}
\{q^{\mu},q^{\upsilon}\}_{D}=0,\{p_{\mu},p_{\nu}\}_{D}=0,\{q^{\mu},p_{\nu
}\}_{D}=\delta_{\nu}^{\mu}.  \label{xp1}
\end{equation}
By use of the equation of motion,%
\begin{equation}
\dot{f}=\left\{ f,H\right\} _{D}.  \label{eqm}
\end{equation}
we obtain those for the positions $\theta$, $\rho$ and the momenta $%
p_{\theta }$, $p_{\rho}$, respectively,%
\begin{align}
\dot{\theta} & \equiv\left\{ \theta,H\right\} _{D}=\frac{p_{\theta}}{%
ma^{2}\cosh^{2}\frac{\rho}{a}},\text{\ \ }\dot{\rho}\equiv\left\{
\rho,H\right\} _{D}=\frac{p_{\rho}}{m\cosh^{2}\frac{\rho}{a}},  \label{xh} \\
\dot{p}_{\theta} & \equiv\left\{ p_{\theta},H\right\} _{D}=0,\text{ \ }\dot{p%
}_{\rho}\equiv\left\{ p_{\rho},H\right\} _{D}=\frac{2}{a}\tanh \frac{\rho}{a}%
H.  \label{ph}
\end{align}
In these calculations (\ref{eqm}) and (\ref{ph}), we used the usual form of
Hamiltonian, $H_{p}\rightarrow H$,%
\begin{equation}
H=\frac{1}{2ma^{2}\cosh^{2}\frac{\rho}{a}}\left( p_{\theta}^{2}+a^{2}p_{\rho
}^{2}\right) .
\end{equation}

So far, the classical mechanics for the motion on the catenoid is complete
and coherent in itself.\ 

\subsection{Quantum mechanical treatment}

In quantum mechanics, we assume that the Hamiltonian takes the following
general form,%
\begin{align}
H& =-\frac{\hbar ^{2}}{2m}\left[ \nabla ^{2}+\left( \alpha M^{2}-\beta
K\right) \right]  \notag \\
& =-\frac{\hbar ^{2}}{2m}[\frac{1}{a^{2}\cosh ^{2}\frac{\rho }{a}}\left( 
\frac{\partial ^{2}}{\partial \theta ^{2}}+a^{2}\frac{\partial ^{2}}{%
\partial \rho ^{2}}\right) +\beta \frac{1}{a^{2}\cosh ^{4}\frac{\rho }{a}}],
\label{h}
\end{align}%
where, 
\begin{equation}
M=0,\text{ }K=-\frac{1}{a^{2}\cosh ^{4}\frac{\rho }{a}}.
\end{equation}%
We are ready to construct commutator $[A,B]$ of two variables $A$ and $B$ in
quantum mechanics, which can be straightforward realized by a direct
correspondence of the Dirac's brackets as $\left[ A,B\right] /i\hbar
\rightarrow $ $\{A,B\}_{D}$. From the Dirac's brackets (\ref{xp1}), the
first category of the fundamental commutators between operators $q^{\mu }$
and $p_{\nu }$ are given by,%
\begin{equation}
\lbrack q^{\mu },q^{\nu }]=0,[p_{\mu },p_{\nu }]=0,[q^{\mu },p_{\nu
}]=i\hbar \delta _{\nu }^{\mu }.  \label{xp2}
\end{equation}%
Similarly, we have the\ second category of fundamental commutators between $%
q^{\mu }$ and $H$ from Eq. (\ref{xh}),%
\begin{align}
\left[ \theta ,H\right] & =\frac{i\hbar }{ma^{2}\cosh ^{2}\frac{\rho }{a}}%
p_{\theta },  \label{qxh1} \\
\left[ \rho ,H\right] & =\frac{i\hbar }{m}\frac{1}{2}\left( \frac{1}{\cosh
^{2}\frac{\rho }{a}}p_{\rho }+p_{\rho }\frac{1}{\cosh ^{2}\frac{\rho }{a}}%
\right) .  \label{qxh2}
\end{align}%
On the other, the quantum commutators (\ref{qxh1}) and (\ref{qxh2}) from
Hamiltonian (\ref{h}) give a definite and satisfactory form for the operator 
$p_{\theta }$ and $p_{\rho }$, \cite{pauli} 
\begin{align}
p_{\theta }& =-i\hbar \frac{\partial }{\partial \theta }, \\
p_{\rho }& =-i\hbar \left( \frac{\partial }{\partial \rho }+\frac{1}{a}\tanh 
\frac{\rho }{a}\right) .  \label{cmom}
\end{align}%
Using these operators, we can directly calculate two quantum commutators $%
\left[ p_{\theta },H\right] $ and $\left[ p_{\rho },H\right] $ with quantum
Hamiltonian (\ref{h}), and the results are, respectively,%
\begin{align}
\left[ p_{\theta },H\right] & =0,  \label{qph1} \\
\left[ p_{\rho },H\right] & =i\hbar \left\{ p_{\theta },H\right\} _{D}-\beta 
\frac{i\hbar ^{3}}{ma^{3}\cosh ^{5}\frac{\rho }{a}}\sinh \frac{\rho }{a}.
\label{qph2}
\end{align}%
The first equation (\ref{qph1}) is satisfactory, whereas the second one (\ref%
{qph2})\ is problematic. With a choice of the parameter\ $\beta \neq 0$,
there is a manifest breakdown of the formal algebraic structure. With a
choice of the parameter\ $\beta =0$, the SCQ becomes self-consistent, but
contradicts with the Schr\"{o}dinger theory that predicts $\beta =1$.

\subsection{Remarks}

From the studies in this section, we see that the SCQ for quantum motion on
the catenoid can be consistently established but is contrary to the Schr\"{o}%
dinger theory. We therefore need to invoke an extrinsic examination of the
same problem, as will be done in next section.

\section{Dirac's theory of second-class constraints for a catenoid as a
submanifold}

The surface equation of the catenoid (\ref{rr}) in Cartesian coordinates $%
\left( x,y,z\right) $ is given by,%
\begin{equation}
f\left( \mathbf{x}\right) \equiv x^{2}+y^{2}-a^{2}\cosh^{2}\frac{z}{a}=0.
\end{equation}
In this section, we will also first give the classical mechanics for motion
on the catenoid within Dirac's theory of second-class constraints, and then
turn into quantum mechanics. The obtained momentum and Hamiltonian are all
compatible with the those given by Schr\"{o}dinger theory.

\subsection{Classical mechanical treatment}

The Lagrangian $L$ in the Cartesian coordinate system is,%
\begin{equation}
L=\frac{m}{2}\left( \dot{x}^{2}+\dot{y}^{2}+\dot{z}^{2}\right) -\lambda
f\left( \mathbf{x}\right) .  \label{lagca}
\end{equation}%
The generalized momentum $p_{i}$ $(i=x,y,z)$ and $p_{\lambda }$ canonically
conjugate to variables $x_{i}$ $(x_{1}=x,x_{2}=y,x_{3}=z,)$ and $\lambda $,
are given by, respectively,%
\begin{align}
p_{i}& =\frac{\partial L}{\partial \dot{x}_{i}}=m\dot{x}_{i},(i=1,2,3), \\
p_{\lambda }& =\frac{\partial L}{\partial \dot{\lambda}}=0.  \label{plambca}
\end{align}%
Eq. (\ref{plambca}) represents the primary constraint,%
\begin{equation}
\varphi _{1}\equiv p_{\lambda }\approx 0.  \label{prim2}
\end{equation}%
By the Legendre transformation, the primary Hamiltonian $H_{p}$ is,%
\begin{equation}
H_{p}=\frac{1}{2m}p_{i}^{2}+\lambda f\left( \mathbf{x}\right) +up_{\lambda }.
\end{equation}%
The secondary constraints are determined by successive use of the Poisson
brackets,%
\begin{align}
\varphi _{2}& \equiv \left\{ \varphi _{1},H_{p}\right\}
=-(x^{2}+y^{2}-a^{2}\cosh ^{2}\frac{z}{a})\approx 0,  \label{cmian} \\
\varphi _{3}& \equiv \left\{ \varphi _{2},H_{p}\right\} =\frac{-2\left(
p_{x}x+p_{y}y\right) +ap_{z}\sinh \frac{2z}{a}}{m}\approx 0, \\
\varphi _{4}& \equiv \left\{ \varphi _{3},H_{p}\right\} =\frac{\lambda
(-a^{2}+8\left( x^{2}+y^{2}\right) +a^{2}\cosh \frac{4z}{a})}{2m}-\frac{%
2(p_{x}^{2}+p_{y}^{2}-p_{z}^{2}\cosh \frac{2z}{a})}{m^{2}}\approx 0.
\label{cthree}
\end{align}%
Similarly, the Dirac bracket between two variables $A$ and $B$ is defined by,%
\begin{equation}
\left\{ A,B\right\} _{D}=\left\{ A,B\right\} -\left\{ A,\varphi _{\xi
}\right\} D_{\xi \zeta }^{-1}\left\{ \varphi _{\zeta },B\right\} ,
\end{equation}%
where the $4\times 4$ matrix $D\equiv \left\{ D_{\xi \zeta }\right\} $ whose
elements are defined by $D_{\xi \zeta }\equiv \left\{ \varphi _{\xi
},\varphi _{\zeta }\right\} $ with $\xi ,\zeta =1,2,3,4$ from Eqs. (\ref%
{prim2}) and (\ref{cmian})-(\ref{cthree}). The inverse matrix $D^{-1}$ is
easily carried out,%
\begin{equation}
D^{-1}=\left( 
\begin{array}{cccc}
0 & D_{12}^{-1} & D_{13}^{-1} & D_{14}^{-1} \\ 
-D_{12}^{-1} & 0 & D_{23}^{-1} & 0 \\ 
-D_{13}^{-1} & -D_{23}^{-1} & 0 & 0 \\ 
-D_{14}^{-1} & 0 & 0 & 0%
\end{array}%
\right) ,
\end{equation}%
where,%
\begin{align}
D_{12}^{-1}& =\frac{\left( 12\left( p_{x}^{2}+p_{y}^{2}\right)
+5p_{z}^{2}-4\left( p_{x}^{2}+p_{y}^{2}+2p_{z}^{2}\right) \cosh \frac{2z}{a}%
+3p_{z}^{2}\cosh \frac{4z}{a}\right) }{8a^{4}m\cosh ^{8}\frac{z}{a}}, \\
D_{13}^{-1}& =-\frac{\tanh \frac{z}{a}}{a^{3}\cosh ^{4}\frac{z}{a}}p_{z}, \\
D_{14}^{-1}& =-D_{23}^{-1}=\frac{m}{4a^{2}\cosh ^{4}\frac{z}{a}}.
\end{align}%
Then primary Hamiltonian $H_{p}$ assumes its usual one: $H_{p}\rightarrow H,$%
\begin{equation}
H=\frac{p_{x}^{2}+p_{y}^{2}+p_{z}^{2}}{2m}.  \label{cHP}
\end{equation}%
All fundamental Dirac's brackets are as follows,%
\begin{align}
\{x_{i},x_{j}\}_{D}& =0  \label{cxxca1} \\
\{x_{i},p_{j}\}_{D}& =\delta _{ij}-\frac{1}{a^{2}\cosh ^{4}\frac{z}{a}}%
\kappa _{i}\kappa _{j},  \label{cxpca1} \\
\{p_{i},p_{j}\}_{D}& =-\frac{1}{a^{2}\cosh ^{4}\frac{z}{a}}\left[ \kappa
_{i}p_{j}\left( \delta _{1j}+\delta _{2j}-\delta _{3j}\cosh \frac{2z}{a}%
\right) -\kappa _{j}p_{i}\left( \delta _{1i}+\delta _{2i}-\delta _{3i}\cosh 
\frac{2z}{a}\right) \right] ,  \label{cxhca1} \\
\{x_{i},H\}_{D}& =\frac{p_{i}}{m}=\dot{x}_{i}, \\
\{p_{i},H\}_{D}& =-\frac{1}{ma^{2}\cosh ^{4}\frac{z}{a}}\kappa _{i}\left(
p_{x}^{2}+p_{y}^{2}+p_{z}^{2}\right) +\frac{2}{ma^{2}\cosh ^{2}\frac{z}{a}}%
\kappa _{i}p_{z}^{2}=\dot{p}_{i},  \label{cphca1}
\end{align}%
\ where $\kappa _{i}=x\delta _{i1}+y\delta _{i2}-a\delta _{i3}\sinh
(z/a)\cosh (z/a)$.

\subsection{Quantum mechanical treatment}

Now let us turn to quantum mechanics. The first category of the fundamental
commutators between operators $x_{i}$ and $p_{i}$ are, by quantization of (%
\ref{cxxca1})-(\ref{cxhca1}),%
\begin{align}
\left[ x_{i},x_{j}\right] & =0,\text{ \ }\left[ x_{i},p_{j}\right] =i\hbar
\left( \delta _{ij}-\frac{1}{a^{2}\cosh ^{4}\frac{z}{a}}\kappa _{i}\kappa
_{j}\right) ,  \label{cxx-xp} \\
\left[ p_{i},p_{j}\right] & =-\frac{i\hbar }{a^{2}\cosh ^{4}\frac{z}{a}}%
\left[ \kappa _{i}\left( \delta _{1j}+\delta _{2j}-\delta _{3j}\cosh \frac{2z%
}{a}\right) p_{j}-\kappa _{j}\left( \delta _{1i}+\delta _{2i}-\delta
_{3i}\cosh \frac{2z}{a}\right) p_{i}\right] ,  \label{cppca2}
\end{align}%
There is a family of the momenta $p_{i}$ that are solutions to the Eq. (\ref%
{cppca2}), as explicitly shown in \cite{japan1992}. With these momenta $%
p_{i} $, we completely do not know the correct form of the quantum
Hamiltonian, as suggested by Eq. (\ref{cHP}). It is therefore understandable
that the quantum Hamiltonian would contain arbitrary parameters.

However, with the help of the second category of the fundamental commutators
as $\left[ x_{i},H\right] $ and $\left[ p_{i},H\right] $, we immediately
find that the momentum from following commutator, 
\begin{equation}
\left[ x_{i},H\right] =i\hbar\frac{p_{i}}{m}.  \label{cxhca2}
\end{equation}
They are, respectively, irrespective of the form of the geometric momentum, 
\begin{align}
p_{x} & =-i\hbar\frac{1}{a\cosh\frac{\rho}{a}}\left( -\sin\theta \frac{%
\partial}{\partial\theta}+a\tanh\frac{\rho}{a}\cos\theta\frac{\partial }{%
\partial\rho}\right) , \\
p_{y} & =-i\hbar\frac{1}{a\cosh\frac{\rho}{a}}\left( \cos\theta \frac{%
\partial}{\partial\theta}+a\tanh\frac{\rho}{a}\sin\theta\frac{\partial }{%
\partial\rho}\right) , \\
p_{z} & =-i\hbar\frac{1}{\cosh^{2}\frac{\rho}{a}}\frac{\partial}{\partial\rho%
}.
\end{align}
They are nothing but the geometric momentum (\ref{geom}) on the catenoid.

As to the form of quantum Hamiltonian, we also assume the general form (\ref%
{h}). For the quantum commutators of the operators $p_{x}$, $p_{y}$ and $H$,
we must have from (\ref{cphca1}), 
\begin{equation}
\lbrack p_{i},H]=i\hbar \frac{1}{2}\left( F_{i}+F_{i}^{\dag }\right)
\end{equation}%
where $F^{\dag }$ denotes the Hermitian conjugate of operator $F$, and 
\begin{equation}
F_{i}=-\frac{2}{a^{2}\cosh ^{4}\frac{z}{a}}\kappa _{i}H+\frac{2}{ma^{2}\cosh
^{2}\frac{z}{a}}\kappa _{i}p_{z}^{2},\text{ }\kappa _{i}=x_{i}-a\delta
_{3i}\sinh (\frac{z}{a})\cosh (\frac{z}{a}).
\end{equation}%
We can easily show that the geometric potential with $\beta =1$ is
compatible with the SCQ. For instance, we have, 
\begin{align}
\left[ p_{i},H\right] & =-\frac{i\hbar }{m}\{mH\frac{1}{a^{2}\cosh ^{4}\frac{%
z}{a}}\kappa _{i}+m\frac{1}{a^{2}\cosh ^{4}\frac{z}{a}}\kappa _{i}H  \notag
\\
& -\frac{1}{2}\frac{1}{4}[\frac{2}{ma^{2}\cosh ^{2}\frac{z}{a}}\kappa
_{i}p_{z}^{2}+p_{z}^{2}\frac{2}{ma^{2}\cosh ^{2}\frac{z}{a}}\kappa _{i}] 
\notag \\
& -\frac{1}{2}\frac{3}{4}[\frac{2}{ma^{2}\cosh ^{2}\frac{z}{a}}%
p_{z}^{2}\kappa _{i}+\kappa _{i}p_{z}^{2}\frac{2}{ma^{2}\cosh ^{2}\frac{z}{a}%
}]\}.  \label{cpxyh}
\end{align}%
Unfortunately, this choice is not unique, and we have also, 
\begin{align}
\left[ p_{z},H\right] & =\frac{i\hbar }{m}\{mH\frac{\tanh \frac{z}{a}}{%
a\cosh ^{2}\frac{z}{a}}+m\frac{\tanh \frac{z}{a}}{a\cosh ^{2}\frac{z}{a}}H+%
\frac{3}{4}[\frac{\tanh \frac{z}{a}}{ma}p_{z}^{2}+p_{z}^{2}\frac{\tanh \frac{%
z}{a}}{ma}]  \notag \\
& +\frac{1}{4}[\frac{\tanh \frac{z}{a}}{ma\cosh ^{2}\frac{z}{a}}%
p_{z}^{2}\cosh ^{2}\frac{z}{a}+\cosh ^{2}\frac{z}{a}p_{z}^{2}\frac{\tanh 
\frac{z}{a}}{ma\cosh ^{2}\frac{z}{a}}]\}.  \label{cpzh}
\end{align}%
From Eqs. (\ref{cpxyh}) and (\ref{cpzh}), we can at least conclude that the
geometric potential given by Dirac's canonical quantization is compatible
with Schr\"{o}dinger theory.

\subsection{Remarks}

An examination of the motion on catenoid as\ a submanifold problem in
Dirac's theory of second-class constraints ensures a highly self-consistent
description. This formalism is also compatible with Schr\"{o}dinger one.

\section{Dirac's theory of second-class constraints for a helicoid within
intrinsic geometry}

The helicoid is with two local coordinates $u\in\left( -\infty,+\infty
\right) ,v\in\left( -\infty,+\infty\right) ,$%
\begin{equation}
\mathbf{r}=(u\cos v,u\sin v,rv),  \label{huv}
\end{equation}
In this section, we will first give the classical mechanics for motion on
the helicoid within Dirac's theory of second-class constraints, and then
turn into quantum mechanics. In classical mechanics, the theory appears
nothing surprising, but after transition to quantum mechanics, it becomes
contradictory to itself.

\subsection{Classical mechanical treatment}

The Lagrangian $L$ in the local coordinate system is, 
\begin{equation}
L=\frac{1}{2}m\left( \dot{v}^{2}\left( r^{2}+u^{2}\right) +2r\dot{r}v\dot{v}+%
\dot{r}^{2}v^{2}+\dot{u}^{2}\right) -\lambda (r-a),
\end{equation}%
where $\lambda $ is the Lagrangian multiplier enforcing the constrained of
motion on the surface, nevertheless, we treat the quantity $\lambda $ as an
additional dynamical variable. The Lagrangian is singular because it does
not contain the "velocity" $\dot{\lambda}$. Hence we need Dirac's theory,
which gives the canonical momenta conjugate to $r,u,v$ and $\lambda $ in the
following,%
\begin{align}
p_{r}& =\frac{\partial L}{\partial \dot{r}}=mv(r\dot{v}+v\dot{r}), \\
p_{u}& =\frac{\partial L}{\partial u}=m\dot{u}, \\
p_{v}& =\frac{\partial L}{\partial \dot{\varphi}}=m\left( r^{2}\dot{v}+rv%
\dot{r}+u^{2}\dot{v}\right) , \\
p_{\lambda }& =\frac{\partial L}{\partial \dot{\lambda}}=0.  \label{hplamb}
\end{align}%
Eq. (\ref{hplamb}) represents the primary constraint:%
\begin{equation}
\varphi _{1}\equiv p_{\lambda }\approx 0,  \label{primhel1}
\end{equation}%
The primary Hamiltonian $H_{p}$ is \cite{dirac2},%
\begin{equation}
H_{p}=\frac{\left( r^{2}+u^{2}\right) p_{r}^{2}-2rvp_{r}p_{v}+v^{2}\left(
u^{2}p_{u}^{2}+p_{v}^{2}\right) }{2mu^{2}v^{2}}+\lambda \left( r-a\right)
+up_{\lambda },
\end{equation}%
where $u$ is also a Lagrangian multiplier guaranteeing that this Hamiltonian
is defined on the symplectic manifold. The Poisson bracket of $\left\{
f,g\right\} $ is with $q_{1}=r,q_{2}=u,q_{3}=v$, and $%
p_{1}=p_{r},p_{2}=p_{u},p_{3}=p_{v}$, 
\begin{equation}
\left\{ f,g\right\} \equiv \frac{\partial f}{\partial q_{k}}\frac{\partial g%
}{\partial p_{k}}+\frac{\partial f}{\partial \lambda }\frac{\partial g}{%
\partial p_{\lambda }}-(\frac{\partial f}{\partial p_{k}}\frac{\partial g}{%
\partial q_{k}}+\frac{\partial f}{\partial p_{\lambda }}\frac{\partial g}{%
\partial \lambda }).
\end{equation}%
The complete set of the secondary constraints is, 
\begin{align}
\varphi _{2}& \equiv \left\{ \varphi _{1},H_{p}\right\} =a-r\approx 0,
\label{hra} \\
\varphi _{3}& \equiv \left\{ \varphi _{2},H_{p}\right\} =\frac{\text{$p_{v}$}%
rv-\text{$p_{r}$}\left( r^{2}+u^{2}\right) }{mu^{2}v^{2}}\approx 0,
\label{pruv} \\
\varphi _{4}& \equiv \left\{ \varphi _{3},H_{p}\right\} =\frac{\left(
r^{2}+u^{2}\right) }{mu^{2}v^{2}}\lambda +\frac{2(\text{$p_{v}$}v-\text{$%
p_{r}$}r)\left( \text{$p_{r}$}\left( r^{2}+u^{2}\right) -p_{u}ruv^{2}-\text{$%
p_{v}$}rv\right) }{m^{2}u^{4}v^{4}}\approx 0.  \label{uvlamb}
\end{align}%
Eqs. (\ref{hra}) shows that on the surface of helicoid $r=a$, while Eqs. (%
\ref{uvlamb}) determines the dynamical variable $\lambda $, and by the
conservation condition of the secondary constraint $\varphi _{4}$ (\ref%
{uvlamb}), we can determine the Lagrangian multipliers $u$.

The Dirac bracket instead of the Poisson bracket for two variables $A$ and $%
B $ is defined by,%
\begin{equation}
\left\{ A,B\right\} _{D}\equiv \left\{ A,B\right\} -\left\{ A,\varphi _{\xi
}\right\} C_{\xi \zeta }^{-1}\left\{ \varphi _{\zeta },B\right\} ,
\end{equation}%
where the $4\times 4$ matrix $C\equiv \left\{ C_{\xi \zeta }\right\} $ whose
elements are defined by $C_{\xi \zeta }\equiv \left\{ \varphi _{\xi
},\varphi _{\zeta }\right\} $ with $\xi ,\zeta =1,2,3,4$ from Eqs. (\ref%
{primhel1}) and (\ref{hra})-(\ref{uvlamb}). The inverse matrix $C^{-1}$ is,%
\begin{equation}
C^{-1}=\left\{ 
\begin{array}{cccc}
0 & C_{12}^{-1} & C_{13}^{-1} & C_{14}^{-1} \\ 
-C_{12}^{-1} & 0 & C_{23}^{-1} & 0 \\ 
-C_{13}^{-1} & -C_{23}^{-1} & 0 & 0 \\ 
-C_{14}^{-1} & 0 & 0 & 0%
\end{array}%
\right\} ,
\end{equation}%
where%
\begin{align}
C_{12}^{-1}& =\frac{2\text{$p_{v}$}uv\left( 3a^{4}\text{$p_{u}$}+2a^{2}u(%
\text{$p_{u}$}u+\text{$p_{v}$}v)-\text{$p_{u}$}u^{4}\right) }{m\left(
a^{2}+u^{2}\right) ^{4}}, \\
C_{13}^{-1}& =\frac{2uv\left( a^{2}\text{$p_{u}$}v+\text{$p_{v}$}u\right) }{%
\left( a^{2}+u^{2}\right) ^{2}}, \\
C_{14}^{-1}& =-C_{23}^{-1}=\frac{mu^{2}v^{2}}{a^{2}+u^{2}}.
\end{align}%
Thus, the generalized positions $q^{\mu }$ $(=u,v)$ and momenta $p_{\nu }$
satisfy the following Dirac brackets,%
\begin{equation}
\{q^{\mu },q^{\nu }\}_{D}=0,\{p_{\mu },p_{\nu }\}_{D}=0,\{q^{\mu },p_{\nu
}\}_{D}=\delta _{\nu }^{\mu }.  \label{uvxp1}
\end{equation}%
By use of the equation of motion,%
\begin{equation}
\dot{f}=\left\{ f,H\right\} _{D}.
\end{equation}%
we obtain those for the positions $u$, $v$ and the momenta $p_{u}$, $p_{v}$,
respectively,%
\begin{align}
\dot{u}& \equiv \left\{ u,H\right\} _{D}=\frac{p_{u}}{m},\text{\ \ }\dot{v}%
\equiv \left\{ v,H\right\} _{D}=\frac{p_{v}}{m\left( u^{2}+a^{2}\right) },
\label{uvxh} \\
\dot{p}_{u}& \equiv \left\{ p_{u},H\right\} _{D}=\frac{up_{v}^{2}}{m\left(
a^{2}+u^{2}\right) ^{2}},\text{ \ }\dot{p}_{v}\equiv \left\{ p_{v},H\right\}
_{D}=0.  \label{uvph}
\end{align}%
In these calculations (\ref{uvxh}) and (\ref{uvph}), we in fact need only
the usual form of Hamiltonian, $H_{p}\rightarrow H$,%
\begin{equation}
H=\frac{1}{2m}\left( p_{u}^{2}+\frac{1}{\left( a^{2}+u^{2}\right) }%
p_{v}^{2}\right) .
\end{equation}

So far, the classical mechanics for the motion on the helicoid is complete
and coherent in itself.\ 

\subsection{Quantum mechanical treatment}

In quantum mechanics, we assume that the Hamiltonian takes the following
general form,%
\begin{align}
H& =-\frac{\hbar ^{2}}{2m}\left[ \nabla ^{2}+\left( \alpha M^{2}-\beta
K\right) \right]  \notag \\
& =-\frac{\hbar ^{2}}{2m}\left( \frac{1}{a^{2}+u^{2}}\frac{\partial ^{2}}{%
\partial v^{2}}+\frac{u}{\left( a^{2}+u^{2}\right) }\frac{\partial }{%
\partial u}+\frac{\partial ^{2}}{\partial u^{2}}+\beta \frac{a^{2}}{\left(
a^{2}+u^{2}\right) ^{2}}\right) ,  \label{uvh}
\end{align}%
where, 
\begin{equation}
M=0,\text{ }K=-\frac{a^{2}}{\left( a^{2}+u^{2}\right) ^{2}}.
\end{equation}%
From the Dirac's brackets (\ref{uvxp1}), the first category of the
fundamental commutators between operators $q^{\mu }$ and $p_{\nu }$ are
given by,%
\begin{equation}
\lbrack q^{\mu },q^{\nu }]=0,[p_{\mu },p_{\nu }]=0,[q^{\mu },p_{\nu
}]=i\hbar \delta _{\nu }^{\mu }.  \label{uvxp2}
\end{equation}%
Similarly, we have the\ second category of fundamental commutators between $%
q^{\mu }$ and $H$ from Eq. (\ref{uvxh}),%
\begin{align}
\left[ u,H\right] & =\frac{i\hbar }{m}p_{u},  \label{uvqxh1} \\
\left[ v,H\right] & =i\hbar \frac{p_{v}}{m\left( u^{2}+a^{2}\right) }.
\label{uvqxh2}
\end{align}%
On the other, the quantum commutators (\ref{uvqxh1}) and (\ref{uvqxh2}) from
Hamiltonian (\ref{uvh}) give a definite and satisfactory form for the
operator $p_{u}$ and $p_{v}$, \cite{pauli}%
\begin{align}
p_{u}& =-i\hbar \left( \frac{\partial }{\partial u}+\frac{u}{2\left(
a^{2}+u^{2}\right) }\right) ,  \label{ucm} \\
p_{v}& =-i\hbar \frac{\partial }{\partial v}.  \label{vcm}
\end{align}%
Using these operators, we can directly calculate two quantum commutators $%
\left[ p_{u},H\right] $ and $\left[ p_{v},H\right] $ with quantum
Hamiltonian (\ref{uvh}), and the results are, respectively,%
\begin{align}
\left[ p_{u},H\right] & =i\hbar \left\{ p_{u},H\right\} _{D}-\frac{i\hbar
^{3}u}{4m\left( a^{2}+u^{2}\right) ^{3}}\left[ a^{2}\left( -5+8\beta \right)
+u^{2}\right] ,  \label{puH} \\
\left[ p_{v},H\right] & =0.  \label{pvH}
\end{align}%
The second equation (\ref{pvH}) is satisfactory, whereas the first one (\ref%
{puH})\ is problematic. there is a manifest breakdown of the formal
algebraic structure between $\left\{ p_{u},H\right\} _{D}$ and $\left[
p_{u},H\right] $ no matter what value of the parameter $\beta $ is chosen.

\subsection{Remarks}

From the studies in this section, we see that the generalized Dirac's theory
of second-class constraints for quantum motion on the helicoid can not be
consistently established. We therefore need to invoke an extrinsic
examination of the same problem, as will be done in next section.

\section{Dirac's theory of second-class constraints for a helicoid as a
submanifold}

The surface equation of the helicoid (\ref{huv}) in Cartesian coordinates $%
\left( x,y,z\right) $ is given by,%
\begin{equation}
f\left( \mathbf{x}\right) \equiv z-a\arctan\frac{y}{x}=0.
\end{equation}
In this section, we will also first give the classical mechanics for motion
on the helicoid within Dirac's theory of second-class constraints, and then
turn into quantum mechanics. The obtained momentum and Hamiltonian are all
in agreement with those given by Schr\"{o}dinger theory.

\subsection{Classical mechanical treatment}

The Lagrangian $L$ in the Cartesian coordinate system is,%
\begin{equation}
L=\frac{m}{2}\left( \dot{x}^{2}+\dot{y}^{2}+\dot{z}^{2}\right) -\lambda
f\left( \mathbf{x}\right) .  \label{helicoid l}
\end{equation}%
The generalized momentum $p_{i}$ $(i=x,y,z)$ and $p_{\lambda }$ canonically
conjugate to variables $x_{i}$ $(x_{1}=x,x_{2}=y,x_{3}=z,)$ and $\lambda $,
are given by, respectively,%
\begin{align}
p_{i}& =\frac{\partial L}{\partial \dot{x}_{i}}=m\dot{x}_{i},(i=1,2,3), \\
p_{\lambda }& =\frac{\partial L}{\partial \dot{\lambda}}=0.  \label{heplamb}
\end{align}%
Eq. (\ref{heplamb}) represents the primary constraint,%
\begin{equation}
\varphi _{1}\equiv p_{\lambda }\approx 0.  \label{heprim2}
\end{equation}%
By the Legendre transformation, the primary Hamiltonian $H_{p}$ is,%
\begin{equation}
H_{p}=\frac{1}{2m}p_{i}^{2}+\lambda f\left( \mathbf{x}\right) +up_{\lambda }.
\end{equation}%
The secondary constraints are determined by\ successive use of the Poisson
brackets,%
\begin{align}
\varphi _{2}& \equiv \left\{ \varphi _{1},H_{p}\right\} =-(z-a\arctan \frac{y%
}{x})\approx 0,  \label{uvxyz} \\
\varphi _{3}& \equiv \left\{ \varphi _{2},H_{p}\right\} =\frac{%
a(p_{y}x-p_{x}y)-p_{z}\left( x^{2}+y^{2}\right) }{m\left( x^{2}+y^{2}\right) 
}\approx 0, \\
\varphi _{4}& \equiv \left\{ \varphi _{3},H_{p}\right\} =\frac{\lambda
\left( a^{2}+x^{2}+y^{2}\right) }{m\left( x^{2}+y^{2}\right) }+\frac{%
2a(-p_{y}x+p_{x}y)(p_{x}x+p_{y}y)}{m^{2}\left( x^{2}+y^{2}\right) ^{2}}%
\approx 0.  \label{uvth}
\end{align}%
Similarly, the Dirac bracket between two variables $A$ and $B$ is defined by,%
\begin{equation}
\left\{ A,B\right\} _{D}=\left\{ A,B\right\} -\left\{ A,\varphi _{\xi
}\right\} D_{\xi \zeta }^{-1}\left\{ \varphi _{\zeta },B\right\} ,
\end{equation}%
where the $4\times 4$ matrix $D\equiv \left\{ D_{\xi \zeta }\right\} $ whose
elements are defined by $D_{\xi \zeta }\equiv \left\{ \varphi _{\xi
},\varphi _{\zeta }\right\} $ with $\xi ,\zeta =1,2,3,4$ from Eqs. (\ref%
{heprim2}) and (\ref{uvxyz})-(\ref{uvth}). The inverse matrix $D^{-1}$ is
easily carried out,%
\begin{equation}
D^{-1}=\left( 
\begin{array}{cccc}
0 & D_{12}^{-1} & D_{13}^{-1} & D_{14}^{-1} \\ 
-D_{12}^{-1} & 0 & D_{23}^{-1} & 0 \\ 
-D_{13}^{-1} & -D_{23}^{-1} & 0 & 0 \\ 
-D_{14}^{-1} & 0 & 0 & 0%
\end{array}%
\right) ,
\end{equation}%
where,%
\begin{align}
D_{12}^{-1}& =\frac{4a^{2}(p_{y}x-p_{x}y)^{2}}{m\left( x^{2}+y^{2}\right)
\left( a^{2}+x^{2}+y^{2}\right) ^{2}}, \\
D_{13}^{-1}& =\frac{2a^{2}(p_{x}x+p_{y}y)}{\left( a^{2}+x^{2}+y^{2}\right)
^{2}}, \\
D_{14}^{-1}& =-D_{23}^{-1}=\frac{m\left( x^{2}+y^{2}\right) }{%
a^{2}+x^{2}+y^{2}}.
\end{align}%
Then primary Hamiltonian $H_{p}$ assumes its usual one: $H_{p}\rightarrow H,$%
\begin{equation}
H=\frac{p_{x}^{2}+p_{y}^{2}+p_{z}^{2}}{2m}.  \label{HP}
\end{equation}%
All fundamental Dirac's brackets are as follows,%
\begin{align}
\{x_{i},x_{j}\}_{D}& =0  \label{xxca1uv} \\
\{x_{i},p_{j}\}_{D}& =\delta _{ij}-\frac{a^{2}}{\left( x^{2}+y^{2}\right)
\left( a^{2}+x^{2}+y^{2}\right) }\chi _{i}\chi _{j},  \label{xpca1uv} \\
\{p_{i},p_{j}\}_{D}& =-\frac{a^{2}}{\left( x^{2}+y^{2}\right) \left(
a^{2}+x^{2}+y^{2}\right) }\left\{ \chi _{i}\left( \delta _{1j}p_{y}-\delta
_{2j}p_{x}+\delta _{3j}\frac{2\left( xp_{x}+yp_{y}\right) }{a}\right) \right.
\notag \\
& \left. -\chi _{j}\left( \delta _{1i}p_{y}-\delta _{2i}p_{x}+\delta _{3i}%
\frac{2\left( xp_{x}+yp_{y}\right) }{a}\right) \right\}  \label{uvpp} \\
\{x_{i},H\}_{D}& =\frac{p_{i}}{m}=\dot{x}_{i},  \label{xhca1uv} \\
\{p_{i},H\}_{D}& =-\frac{2a\left( p_{x}x+p_{y}y\right) }{\left(
x^{2}+y^{2}\right) \left( a^{2}+x^{2}+y^{2}\right) }p_{z}\chi _{i}=\dot{p}%
_{i},  \label{phca1}
\end{align}%
\ where $\chi _{i}=\delta _{1i}y-\delta _{2i}x+\delta _{3i}\left(
x^{2}+y^{2}\right) /a$.

\subsection{Quantum mechanical treatment}

Now let us turn to quantum mechanics. The first category of the fundamental
commutators between operators $x_{i}$ and $p_{i}$ are, by quantization of (%
\ref{xxca1uv})-(\ref{uvpp}),%
\begin{align}
\left[ x_{i},x_{j}\right] & =0,\text{ \ }\left[ x_{i},p_{j}\right] =i\hbar
\left( \delta _{ij}-\frac{a^{2}}{\left( x^{2}+y^{2}\right) \left(
a^{2}+x^{2}+y^{2}\right) }\chi _{i}\chi _{j}\right) ,  \label{xx-xpuv} \\
\left[ p_{i},p_{j}\right] & =i\hbar \{p_{i},p_{j}\}_{D}.  \label{ppca2uv}
\end{align}%
However, with the help of the second category of the fundamental commutators
as $\left[ x_{i},H\right] $ and $\left[ p_{i},H\right] $, we immediately
find that the momentum from following commutator, 
\begin{equation}
\left[ x_{i},H\right] =i\hbar \frac{p_{i}}{m}.  \label{xhca2}
\end{equation}%
They are, respectively, irrespective of the form of the geometric potential, 
\begin{align}
p_{x}& =-i\hbar \left( \cos v\frac{\partial }{\partial u}-\frac{u\sin v}{%
a^{2}+u^{2}}\frac{\partial }{\partial v}\right) , \\
p_{y}& =-i\hbar \left( \sin v\frac{\partial }{\partial u}+\frac{u\cos v}{%
a^{2}+u^{2}}\frac{\partial }{\partial v}\right) , \\
p_{z}& =-i\hbar \frac{a}{a^{2}+u^{2}}\frac{\partial }{\partial v}.
\end{align}%
They are nothing but the geometric momentum (\ref{geom}) on the helicoid.

As to the form of quantum Hamiltonian, we also assume the general form (\ref%
{uvh}). For the quantum commutators of the operators $p_{x}$, $p_{y}$, $%
p_{z} $ and $H$, we must have from (\ref{phca1}), 
\begin{equation}
\lbrack p_{i},H]=i\hbar \frac{1}{2}\left( F_{i}+F_{i}^{\dag }\right)
\end{equation}%
where 
\begin{equation}
F_{i}=-\frac{2a\left( p_{x}x+p_{y}y\right) }{\left( x^{2}+y^{2}\right)
\left( a^{2}+x^{2}+y^{2}\right) }p_{z}\chi _{i},\text{ }\chi _{i}=\delta
_{1i}y-\delta _{2i}x+\delta _{3i}\frac{\left( x^{2}+y^{2}\right) }{a}.
\end{equation}%
We can easily show that the geometric potential with $\beta =1$ is
compatible with the SCQ. For instance, we have,%
\begin{align}
\left[ p_{i},H\right] & =i\hbar \left\{ -\alpha _{1}\left[ xg_{i}\frac{1}{2}%
\left( p_{x}p_{z}+p_{z}p_{x}\right) +(x\rightleftharpoons y,p_{x}\rightarrow
p_{y})\right] \right.  \notag \\
& -\alpha _{2}\left[ \frac{1}{2}\left( p_{x}p_{z}+p_{z}p_{x}\right)
xg_{i}+(x\rightleftharpoons y,p_{x}\rightarrow p_{y})\right]  \notag \\
& -\left. \alpha _{3}\left[ \frac{1}{2}\left(
p_{x}xg_{i}p_{z}+p_{z}xg_{i}p_{x}\right) +(x\rightleftharpoons
y,p_{x}\rightarrow p_{y})\right] \right\}
\end{align}%
where $\alpha _{k}$, $(k=1,2,3)$ are three real parameters satisfying $\sum
\alpha _{k}=1$, $g_{i}=2a\chi _{i}/(m\left( x^{2}+y^{2}\right) \left(
a^{2}+x^{2}+y^{2}\right) )$. In comparison of both sides of the this
equation, we find $\beta $ and three real parameters $\alpha _{k}$ are
freely to be specified, 
\begin{equation}
\alpha _{1}=\alpha _{2}=\frac{1}{2}-\frac{\alpha _{3}}{2},\beta =-\frac{3}{4}%
(1-\alpha _{3}).
\end{equation}%
when the free parameter $\alpha _{3}$ is defined as $-1/3$, we can find $%
\beta =1$ in (\ref{uvh}), the geometric potential given by Dirac formalism
matches with Schr\"{o}dinger theory.

\subsection{Remarks}

An examination of the motion on helicoid as\ a submanifold problem in
Dirac's theory of second-class constraints ensures a self-consistent
description.

\section{ Discussions and conclusions}

A revisit of relation between Cartesian coordinates and canonical
quantization scheme for quantum motions on catenoid\ and helicoid is done
extensively. Within the intrinsic geometry, the generalized momenta can all
be obtained within the SCQ, but the geometric potentials can hardly be
self-consistently formulated into the theory in general. In contrast,
canonical quantization in three-dimensional flat space in which the surface
is embedded is successful, and we have not only the geometric momenta, but
also the geometric potentials that can be after all obtained with help of
the rearrangement of the operator-ordering. We can safely conclude that
intrinsic geometry does not in general offer a framework for quantum
mechanics to be satisfactorily formulated. 

This study is compatible with the Dirac's insightful remark, implying that
the preferable Cartesian coordinate system has fundamental importance in
passing from the classical Hamiltonian to its quantum mechanical form. For
particles move on the curved manifold, we have to avoid the intrinsic
geometry, and search for higher-dimensional flat space the manifold is
embedded.

\section*{Acknowledgments}

This work is financially supported by National Natural Science Foundation of
China under Grant No. 11175063.

\end{document}